\documentclass[%
reprint,
superscriptaddress,
 amsmath,amssymb,
 aps,
pra,
floatfix,
]{revtex4-2}

\usepackage[section]{placeins}
\usepackage{floatrow}
\usepackage{physics} %for physics (vector) added by NV
\usepackage{chemformula} %for chemistry added by NV
\usepackage{graphicx}% Include figure files
\usepackage{dcolumn}% Align table columns on decimal point
\usepackage{bm}% bold math
\usepackage{amsmath}
\usepackage{ulem} 
\usepackage{lipsum}% http://ctan.org/pkg/lipsum
\usepackage{color, hyperref}
\usepackage{units}
\usepackage[all]{hypcap}
\usepackage{chemfig}
\usepackage{tikz}
\usepackage[version=4]{mhchem}
\usepackage{graphicx}
\usepackage{epstopdf}
\usepackage{xcite}
\usepackage{multibib}

\begin{document}
\preprint{APS/123-QED}

%\title{Ultrafast Nuclear Dynamics in Double-Core-Hole States of  Water Molecules}% Force line breaks with \\
\title{Ultrafast Nuclear Dynamics in Double-Core-Ionized Water Molecules}% Force line breaks with \\

\author{Iyas~Ismail}%
\thanks{These authors contributed equally to this work}
\email{iyas.ismail@sorbonne-universite.fr}
\affiliation{%
Sorbonne Université, CNRS, Laboratoire de Chimie Physique-Matière et Rayonnement, LCPMR, F-75005 Paris Cedex 05, France
}%

\author{Ludger~Inhester}%
\thanks{These authors contributed equally to this work}

\email{ludger.inhester@cfel.de}
\affiliation{%
	Center for Free-Electron Laser Science CFEL, Deutsches Elektronen-Synchrotron DESY, Notkestr. 85, 22607 Hamburg, Germany
}%

\author{Tatiana~Marchenko}%
\affiliation{%
	Sorbonne Université, CNRS, Laboratoire de Chimie Physique-Matière et Rayonnement, LCPMR, F-75005 Paris Cedex 05, France
}%

\author{Florian~Trinter}%
\affiliation{%
	Institut für Kernphysik, Goethe-Universität Frankfurt, Max-von-Laue-Straße 1, 60438 Frankfurt am Main, Germany}%

\affiliation{%
	Molecular Physics, Fritz-Haber-Institut der Max-Planck-Gesellschaft, Faradayweg 4-6, 14195 Berlin, Germany}%

\author{Abhishek~Verma}%
\affiliation{%
	Sorbonne Université, CNRS, Laboratoire de Chimie Physique-Matière et Rayonnement, LCPMR, F-75005 Paris Cedex 05, France
}%

\author{Alberto~De~Fanis}%
\affiliation{%
	European XFEL, Holzkoppel 4, 22869 Schenefeld, Germany
}%

\author{Anthony~Ferté}%
\affiliation{%
Sorbonne Université, CNRS, Laboratoire de Chimie Physique-Matière et Rayonnement, LCPMR, F-75005 Paris Cedex 05, France
}%

\author{Daniel~E.~Rivas}%
\affiliation{%
	European XFEL, Holzkoppel 4, 22869 Schenefeld, Germany
}%

\author{Dawei~Peng}%
\affiliation{%
	Sorbonne Université, CNRS, Laboratoire de Chimie Physique-Matière et Rayonnement, LCPMR, F-75005 Paris Cedex 05, France
}%
\affiliation{%
	European XFEL, Holzkoppel 4, 22869 Schenefeld, Germany
}%

\author{Dimitris~Koulentianos}%

\affiliation{%
	Chemical Sciences and Engineering Division, Argonne National Laboratory, 9700 South Cass Avenue, Lemont, Illinois 60439, USA}%

\author{Edwin~Kukk}%
\affiliation{%
	Sorbonne Université, CNRS, Laboratoire de Chimie Physique-Matière et Rayonnement, LCPMR, F-75005 Paris Cedex 05, France
}%
\affiliation{%
	Department of Physics and Astronomy, University of Turku, FI-20014 Turku, Finland}%
 
\author{Francis~Penent}%
\affiliation{%
Sorbonne Université, CNRS, Laboratoire de Chimie Physique-Matière et Rayonnement, LCPMR, F-75005 Paris Cedex 05, France
}%

\author{Gilles~Doumy}%
\affiliation{%
	Chemical Sciences and Engineering Division, Argonne National Laboratory, 9700 South Cass Avenue, Lemont, Illinois 60439, USA}%
 
\author{Giuseppe~Sansone}%
\affiliation{%
Institute of Physics, University of Freiburg, Hermann-Herder-Str. 3, 79104 Freiburg, Germany}%

\author{John~D.~Bozek}%
\affiliation{%
	Synchrotron SOLEIL, L'Orme des Merisiers, Saint-Aubin, F-91192 Gif-sur-Yvette Cedex, France
}%

\author{Kai~Li}%
\affiliation{%
	Chemical Sciences and Engineering Division, Argonne National Laboratory, 9700 South Cass Avenue, Lemont, Illinois 60439, USA}%

\author{Linda~Young}%
\affiliation{%
	Chemical Sciences and Engineering Division, Argonne National Laboratory, 9700 South Cass Avenue, Lemont, Illinois 60439, USA}%
\affiliation{%
	Department of Physics and James Franck Institute, The University of Chicago, Chicago, Illinois, USA
}

\author{Markus~Ilchen}%
\affiliation{%
	European XFEL, Holzkoppel 4, 22869 Schenefeld, Germany
}%

\author{Maria~Novella~Piancastelli}%
\affiliation{%
	Sorbonne Université, CNRS, Laboratoire de Chimie Physique-Matière et Rayonnement, LCPMR, F-75005 Paris Cedex 05, France
}%
\author{Michael~Meyer}%
\affiliation{%
	European XFEL, Holzkoppel 4, 22869 Schenefeld, Germany
}%

\author{Nicolas~Velasquez}%
\affiliation{%
	Sorbonne Université, CNRS, Laboratoire de Chimie Physique-Matière et Rayonnement, LCPMR, F-75005 Paris Cedex 05, France
}%

\author{Oksana~Travnikova}%
\affiliation{%
	Sorbonne Université, CNRS, Laboratoire de Chimie Physique-Matière et Rayonnement, LCPMR, F-75005 Paris Cedex 05, France
}%

\author{Rebecca~Boll}%
\affiliation{%
	European XFEL, Holzkoppel 4, 22869 Schenefeld, Germany
}%

\author{Renaud~Guillemin}%
\affiliation{%
Sorbonne Université, CNRS, Laboratoire de Chimie Physique-Matière et Rayonnement, LCPMR, F-75005 Paris Cedex 05, France
}%

\author{Reinhard~Dörner}%
\affiliation{%
	Institut für Kernphysik, Goethe-Universität Frankfurt, Max-von-Laue-Straße 1, 60438 Frankfurt am Main, Germany}%

\author{Richard~Taïeb}%
\affiliation{%
	Sorbonne Université, CNRS, Laboratoire de Chimie Physique-Matière et Rayonnement, LCPMR, F-75005 Paris Cedex 05, France
}%

\author{Simon~Dold}%
\affiliation{%
	European XFEL, Holzkoppel 4, 22869 Schenefeld, Germany
}%

\author{Stéphane~Carniato}%
\affiliation{%
	Sorbonne Université, CNRS, Laboratoire de Chimie Physique-Matière et Rayonnement, LCPMR, F-75005 Paris Cedex 05, France
}

\author{Thomas~M.~Baumann}%
\affiliation{%
	European XFEL, Holzkoppel 4, 22869 Schenefeld, Germany
}%

\author{Tommaso~Mazza}%
\affiliation{%
	European XFEL, Holzkoppel 4, 22869 Schenefeld, Germany
}%

\author{Yevheniy~Ovcharenko}%
\affiliation{%
	European XFEL, Holzkoppel 4, 22869 Schenefeld, Germany
}%

\author{Ralph~Püttner}%
\affiliation{%
Fachbereich Physik, Freie Universität Berlin, Arnimallee 14 D-14195 Berlin, Germany
}%
\author{Marc~Simon}%
%\email{marc.simon@sorbonne-universite.fr}
\affiliation{%
Sorbonne Université, CNRS, Laboratoire de Chimie Physique-Matière et Rayonnement, LCPMR, F-75005 Paris Cedex 05, France
}%

\date{\today}% It is always \today, today,

	\begin{abstract}
Double-core-hole (DCH) states in isolated water and heavy water molecules, resulting from the sequential absorption of two x-ray photons, have been investigated. A comparison of the subsequent Auger emission spectra from the two isotopes provides direct evidence of ultrafast nuclear motion during the $\unit[1.5]{fs}$ lifetime of these DCH states. Our numerical results align well with the experimental data, providing for various DCH states an in-depth study of the dynamics  responsible of the observed isotope effect.
	
\end{abstract}

\maketitle

%\tableofcontents
\section{\label{sec:intro}Introduction}
Double-core-hole (DCH) states refer to electronic states with two vacancies in the core level.
Spectroscopy of DCH states stands out as a highly promising tool, showing a remarkably enhanced sensitivity to the chemical environment when compared to single-core-hole (SCH) spectroscopy \cite{cederbaum1986double, Santra, salen_2012_experimental, tashiro_2010_molecular, berrah2011double, lablanquie2011evidence, koulentianos2018double, takahashi_2018_theoretical, Goldsztejn, PuttnerPhysRevLett.114.093001}. 
Moreover, specifically for double vacancies in the K-shell, the significantly lower lifetime of these states, compared to SCH states \cite{inhester2013core, Goldsztejn, marchenko2018ultrafast}, positions this emerging  spectroscopy as a powerful femtosecond probe, enabling  the tracking of nuclear dynamics occurring during the first (sub)femtoseconds of the interaction with the light.

%Moreover, the brief lifetime of these states, compared to SCH states \cite{inhester2013core, Goldsztejn, marchenko2018ultrafast}, {\color{red} along with their repulsive nature~\cite{Travnikova},  leads to ultrafast dissociation. This }\textbf{\color{red} this modification does not make sense: In what sense does the repulsive nature of DCH help to track fast nuclear dynamics? I understand that one can consider DCH states as a probe with a short time-window that is sensitive to dynamics. But now you bring up that the probe itself induced dynamcis: How does that help to track them? This requires some elaboration}positions this emerging spectroscopy as a powerful femtosecond probe, allowing for the tracking of nuclear dynamics occurring during the first (sub)femtoseconds of the interaction with the light.

Core-hole and double-core-hole states have received considerable attention due to the fact that these hollow electronic configurations have a temporarily reduced x-ray absorption cross section, leading to an effect termed x-ray-induced transparency or frustrated absorption~\cite{hoener_ultraintense_2010, young_femtosecond_2010}.
%which implies that scattering cross section are compared relatively to ionization cross section.
This becomes relevant for prospective single-molecule diffractive imaging, 
where radiation damage can potentially be suppressed through sufficiently short x-ray pulses~\cite{berrah2011double, son_breakdown_2020}.

Double vacancies in the core level can be produced through one-photon absorption using, e.g., a synchrotron-light source~\cite{marchenko2018ultrafast,koulentianos2018double, Goldsztejn,PuttnerPhysRevLett.114.093001, nakano2013near, mucke2013formation, eland2010double, lablanquie2011properties, lablanquie2011evidence, linusson2011structure, HikosakaPhysRevLett.98.183002,
KanterPhysRevLett.83.508,marchenko_2020_single,koulentianos_2021_CO,koulentianos_2022_using,mailhiot_2023_multielectron}. In this case, DCH states appear as satellites of the one-photon core ionization~\cite{carniato2015single} and hence are referred to as DCH hypersatellites. Furthermore,
the use of highly intense light sources such as x-ray free-electron laser (XFEL) facilities, enable the creation of DCH states through sequential two-photon absorption~\cite{hoener_ultraintense_2010, FangPhysRevLett.105.083005, CryanPhysRevLett.105.083004, tashiro_2010_molecular, young_femtosecond_2010, DoumyPhysRevLett.106.083002,  salen_2012_experimental, tamasaku_2013_double, larsson2013double, zhaunerchyk2013using, frasinski2013dynamics, koulentianos2020high, KastirkePhysRevLett.125.163201, frasinski2013dynamics,Mazza, li_2021_electron, Ismail_PRL, liu2024transient}. For that purpose, the absorption of the second photon has to occur before the decay of the SCH state.
Previous researches have explored a large variety of such states.
Specifically for vacancies in the K-shell, states have been investigated, where either both core-level electrons have been ejected to the continuum (K$^{-2}$ state) or one electron has been ejected to the continuum while the other one is excited to a vacant orbital V (K$^{-2}$V states), or both core-level electrons have been excited to vacant orbitals (K$^{-2}$VV$^\prime$ states). 

%In general, K$^{-2}$-DCH states have a significantly lower lifetime than K$^{-1}$-SCH states~\cite{inhester2013core}. said before
For low-$Z$ elements, K-shell holes decay predominantly non-radiatively via Auger decay.
In this process, an electron from a valence level fills one of the core holes, while another electron is emitted to the continuum. 
For DCH states in the K-shell, this Auger electron has significantly higher kinetic energy compared to the Auger electron created from SCH states. 
This difference in kinetic energy allows for the separation of the signal originating from DCH states from the signal of the SCH states.
The emitted Auger electron is thereby a clear probe of DCH states.

%{\color{red}
In order to characterize the dynamics of DCH states, which is potentially relevant for X-ray diffractive imaging, and to exploit the chemical sensitivity of their spectroscopic signals, it is crucial to understand the electronic spectra of DCH states. 
Earlier studies demonstrated the presence of strongly dissociative potential energy surfaces (PES) in DCH states of \ce{HCl} and \ce{CH3I} molecules~\cite{Travnikova,MarchenkoPhysRevLett.119.133001}. 
For water, a theoretical study by Inhester et al.~\cite{inhester_auger_2012} showed that the produced Auger spectrum is considerably influenced by the involved steep PES.
This prediction was corroborated by a recent study~\cite{marchenko2018ultrafast}, conducted by some of the authors, which revealed an indication of ultrafast proton motion within the very short lifetime (about \unit[1.5]{fs}~\cite{inhester_auger_2012}) of the DCH states in water molecules, leaving significant fingerprints in the Auger emission spectrum.
Furthermore, within the same study, theoretical predictions indicated significant variations in the dynamic response depending on the specific populated DCH states, namely K$^{-2}$ or K$^{-2}$V. Notably, some K$^{-2}$V states turned out to be more dissociative than the bare K$^{-2}$ state.

This theoretical finding could not be directly confirmed through experimental observation because the photon intensity provided by the synchrotron radiation source only allowed the simultaneous population of multiple DCH states via one-photon shake-up and shake-off processes. 
Consequently, only an overall Auger spectrum containing contributions from several DCH (K$^{-2}$, K$^{-2}$V) states could be measured.

In this paper, we present a more direct measurement of the Auger emission from DCH states using an XFEL light source, combined with a comprehensive theoretical investigation of the ultrafast motion following the generation of \textit{specific} DCH states in water molecules.
In particular, by employing x-ray light of sufficient intensity, we study DCH states produced via sequential absorption of two photons.
By tuning the employed photon energy, we explore the spectrum variations of the emitted Auger electron  when addressing regimes leading predominantly 
to K$^{-2}$ states via sequential core-shell photoionization or to K$^{-2}$V states via subsequent core-shell photoionization and core-shell photoexcitation.
This enables distinct investigations of the subsequent dynamics associated with each of these states. 
Moreover, by comparing Auger spectra obtained from isolated water (\ce{H2O}) and heavy water (\ce{D2O}) molecules, we provide now an unequivocal evidence of ultrafast dynamics occurring during the \unit[1.5]{fs} lifetime of the DCH state.

The outline of this article is as follows:
Section~\ref{sec:Experiment} describes the experimental setup,
Sec.~\ref{sec:Theory} describes the calculation methodology. 
The resulting measured and calculated Auger spectra are discussed in Sec.~\ref{sec:Results} and, in Sec.~\ref{sec:Conclusions}, we draw final conclusions.

\begin{figure}[h]
	\includegraphics[angle=0,origin=c,width=0.9\linewidth]{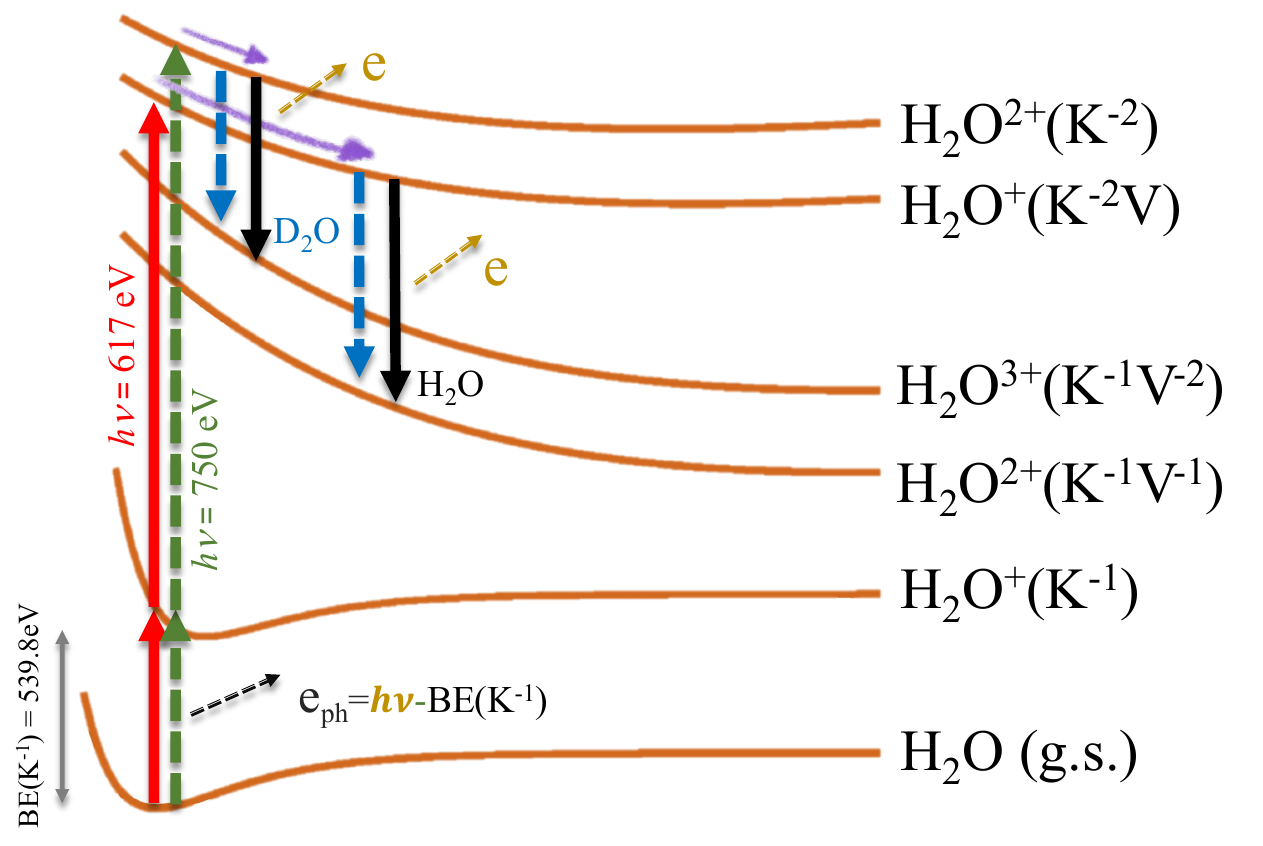}    
	\caption{
		Schematic illustration of the potential energy surfaces of K$^{-2}$ and K$^{-2}$V  along the dissociation coordinate. 
        Using two photons of \unit[750]{eV} energy, the K$^{-2}$ state is produced through sequential ionization: H$_2$O (g.s.) $\rightarrow$ H$_2$O$^+$ (K$^{-1}$) $\rightarrow$ H$_2$O$^{2+}$ (K$^{-2}$).
        Nuclear motion occurs along the dissociative K$^{-2}$  potential energy surface.
        Over the \unit[1.5]{fs} lifetime, the protons explore a greater distance  along the dissociation coordinate than the deuterons, leading to noticeable differences in the Auger electron energy, which are for water (black arrow) larger than for heavy water (blue arrow). 
        With photons of \unit[617]{eV} energy, the K$^{-2}$V state is populated via the following excitation scheme: H$_2$O (g.s.) $\rightarrow$ H$_2$O$^+$ (K$^{-1}$) $\rightarrow$ H$_2$O$^{+}$ (K$^{-2}$V).
        In this case, a similar nuclear motion to that of the K$^{-2}$ state arises.}
		\label{fig:process}
\end{figure}

%Moreover, a clear confirmation of ultrafast hydrogen motion and its impact on the emitted Auger electrons 

%This limitation can be effectively overcome by producing the DCH states with two-photon absorption (Fig~\ref{fig:process}): The first photon ionizes one $1s$ electron and the second photon excites the remaining $1s$ electron into an unoccupied orbital (V).

%With careful tuning of the photon energy, selective population of either specific monocationic K$^{-2}$V states or dicationic K$^{-2}$ states can be achieved. 

%It is important to note that achieving the absorption of two photons within the short lifetime of the $1s^{-1}$ hole requires the use of highly intense and extremely short light pulses, which are uniquely available at the XFEL source facilities.

%We have found that the molecular motion speed and symmetry depend significantly on the populated DCH states. This finding provides a distinctive tool of manipulating molecular fragmentation within a subfemtosecond timescale by tuning the photon energy to populate chosen DCH states.

\section{Experiment}\label{sec:Experiment}
The experiment was conducted using the atomic-like quantum systems (AQS) end-station at the Small Quantum System (SQS) instrument located at the SASE3 undulator of the European XFEL~\cite{app7060592, Mazza:yi5133,decking:2020bct}. The XFEL beam, with a pulse energy of approximately~\unit[4]{mJ}, delivered 30 pulses per train at a repetition rate of \unit[1.1]{MHz} within the train, while the trains operated at a repetition rate of \unit[10]{Hz}. 
The XFEL beam pulses, with a duration of about \unit[25]{fs} and a bandwidth of roughly 6 to \unit[7]{eV} for photon energies in the range from 600 to \unit[750]{eV}, were focused to approximately 1.5 $\times$ 1.5 $\mu$m$^{2}$.
Electrons were detected with a time-of-flight (TOF) spectrometer installed, facing the AQS chamber ionization region, where the x-ray focus was optimized.
The spectrometer was oriented at the magic angle (54.7$^\circ$) with respect to the horizontal linear polarization of the photon beam.
A retardation potential of \unit[490]{V} was applied to improve the energy resolution of the spectrometer for the measured Auger electrons. The energy resolution was about $\unit[900]{meV}$ for kinetic energy around $\unit[550]{eV}$.
The electrons were detected using commercially available microchannel plate (MCP) detectors.
More details about the TOF spectrometer can be found elsewhere~\cite{DeFanis22}.
Water molecules H$_{2}$O and  D$_{2}$O  were introduced using a bubbler system connected to the ultrahigh-vacuum interaction chamber resulting in a pressure of about $\unit[4.6\times 10^{-8}]{mbar}$ in the experimental chamber under gas load.
The TOF spectrometer was energy-calibrated using the H$_{2}$O outer-valence photoelectron lines.
The photon energy was calibrated with the K$^{-2}$ states of water~\cite{marchenko2018ultrafast}.
Additionally, the TOF transmission curves were derived by normalizing the yield of the H$_{2}$O photoelectron lines.

%Starting from ground state (gs), i.e., gs $\rightarrow$ K$^{-1} \rightarrow$ K$^{-2}$V or gs $\rightarrow$ K$^{-1} \rightarrow$ K$^{-2}$. Photon energy was tuned to \unit[617]{eV}  and \unit[750]{eV} allowing the   population of K$^{-2}$V and K$^{-2}$ states, respectively, as outlined in our resent study~\cite{Ismail_PRL}. 
\section{Theoretical Calculations} \label{sec:Theory}
In order to calculate the DCH Auger spectra of water, we employed the \textsc{xmolecule} electronic-structure toolkit (version~3869)~\cite{hao_efficient_2015, inhester_xray_2016}.
Specifically, we computed double-core-ionized states (K$^{-2}$ and various K$^{-2}$V configurations) using the restricted Hartree Fock method. Convergence to the desired states was achieved based on 
the maximum-overlap method~\cite{bagus_1965_self,gilbert_self-consistent_2008}.
All calculations were performed using the aug-cc-pVTZ basis set~\cite{dunning1989a, kendall1992a}.
With the orbitals optimized for the initial, double-core-ionized state, we compute the final electronic states using configuration interaction. 
All direct final Auger configurations (one core hole and two valence holes)
as well as configurations with additional excitation into previously unoccupied (virtual) orbitals 
(one core hole, three valence holes, and one virtual electron) were taken into account for the configurational expansion of the final states.

The Auger amplitudes were calculated using the one-center approximation~\cite{siegbahn_auger_1975} that treats the Auger process as a largely intra-atomic transition and approximates the molecular continuum wave functions with the corresponding atomic continuum wave functions. In \textsc{xmolecule}, atomic continuum matrix elements from atomic calculations using \textsc{xatom}~\cite{jurek_xmdyn_2016} are used.

The effects of nuclear dynamics in the spectrum were incorporated following Refs.~\onlinecite{inhester_auger_2012, marchenko2018ultrafast}. 
A set of 100 molecular dynamics (MD) trajectories were propagated on the respective double core ionized state with a time step of $\unit[0.1]{fs}$ for a total time of $\unit[20]{fs}$. 
These trajectories started from initial conditions sampled from the neutral ground-state Wigner distribution.
For each time step of the MD trajectories, instantaneous Auger spectra $T(E,t)$ were calculated, where for each transition a Lorentzian line shape was employed.
The total Auger spectrum was then compiled using~\cite{marchenko2018ultrafast}
\begin{equation}
T(E) = \int dt T(E,t) e^{-\Gamma t},
\end{equation}
where $\Gamma$ is the calculated reciprocal lifetime of the double-core-hole configuration.
For the final spectrum, we also take into account a further convolution with a Gaussian function with a full width at half maximum of $\unit[900]{meV}$ to account for the finite resolution in the experiment.
For a better comparison with the experiment, the calculated Auger spectra were further shifted by $\unit[5]{eV}$ to lower energies.
This shift compensates for an imbalance between the initial and final electronic states that occur due to the orbital optimization for the initial electronic state and the inclusion of several valence-to-virtual excitations for the final electronic states.
In addition, it compensates for relativistic effects in the oxygen core level, that are not considered in the calculation.

\section{RESULTS AND DISCUSSION}\label{sec:Results}

Figure~\ref{fig:process} sketches the PESs that are populated in the experiment via subsequent core-ionization and core-excitation steps.
Whereas there is little dynamics involved in the first core-ionization step~\cite{inhester_auger_2012}, the second core-ionization or core-excitation step induces ultrafast dissociation of the molecule.
The emission of an Auger electron brings the molecule energetically down either to the dicationic or the tricationic states.
Because these states involve a larger charge in the molecule, the PESs along the dissociation coordinate become in general steeper.
Because of this differential gradient of the PESs along the dissociation coordinate,
the emission energy of the Auger electron shifts to higher energies, the further the dissociative dynamics in the DCH state proceeds.
We note that the scheme presented in Fig.~\ref{fig:process} is able 
to map out dynamics on a timescale much faster than the pulse duration, which is for the current experiment about $\unit[25]{fs}$.
This is enabled by focusing on features specifically related to DCH states that have a very short lifetime, and the fact that no essential dynamics occurs on the preceding SCH state~\cite{inhester_auger_2012}.

\begin{figure}[h]
\includegraphics[angle=0,origin=c,width=0.95\linewidth]{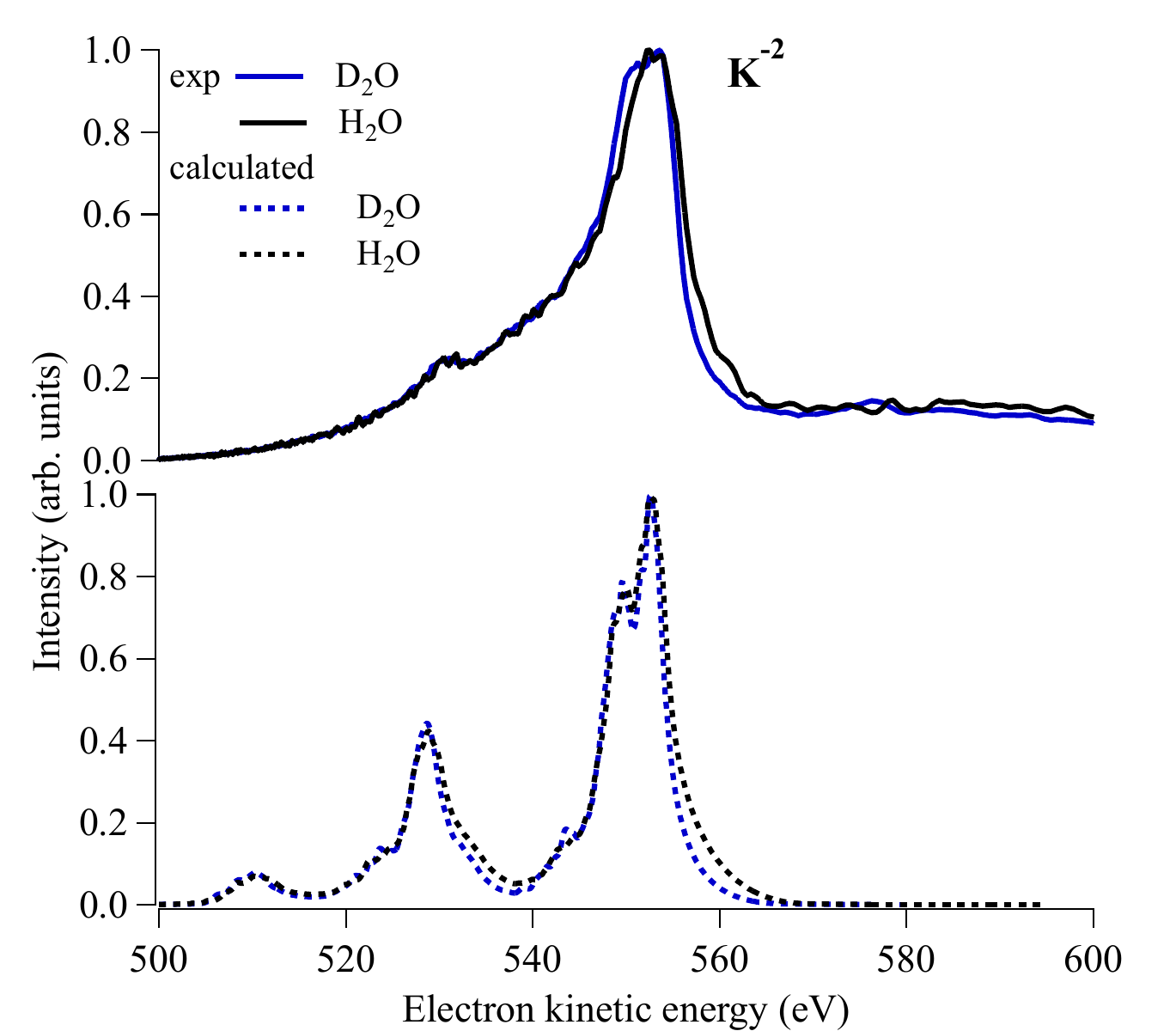}    
	\caption{Comparison of experimental  Auger spectra measured with a photon energy of \unit[750]{eV} (solid lines) with calculated  Auger spectra for the $\mathrm{K}^{-2}$ state (dashed lines). Black lines show spectra for water, blue lines show spectra for heavy water. The spectra are normalized to have similar peak heights.
	\label{fig:exp1}}
\end{figure}

The experimental Auger spectra, measured at a photon energy of \unit[750]{eV} for both water and heavy water, are shown in Fig.~\ref{fig:exp1} along with the calculated spectra for comparison. 
At this photon energy, the x-ray pulse can core-ionize the water molecule in sequential steps, producing mostly the bare $\mathrm{K}^{-2}$ state. 
The difference between water and heavy water can be seen mainly at $\unit[560]{eV}$, on the high-energy side of the dominant Auger peak. 
One can see that the tail to higher energies is much more pronounced for water than for heavy water.
This trend is in good agreement with the calculated Auger spectra. 

Overall, the calculation reproduces all the characteristic features of the experimental Auger spectrum.
However, the experimental spectra appear broader than the calculated ones. Specifically, the lower peak at $\unit[530]{eV}$ is almost completely covered within a broad Auger signal, whereas the calculated Auger spectrum exhibits much more pronounced features.
The respective final states associated with this feature involve highly excited configurations with a core hole, an inner-valence hole, and an outer-valence hole~\cite{inhester_auger_2012}. 
We speculate that the strong repulsion on the respective final-state surfaces is somehow underestimated by the employed classical-trajectory approach, and the computed broadening effect of these Auger lines might therefore be too low.
%The clear difference in \ce{H2O} and \ce{D2O} Auger spectrum clearly establishes that the extended tail is an effect of the ultrafast dynamics.

\begin{figure}[h]
	\includegraphics[angle=0,origin=c,width=0.95\linewidth]{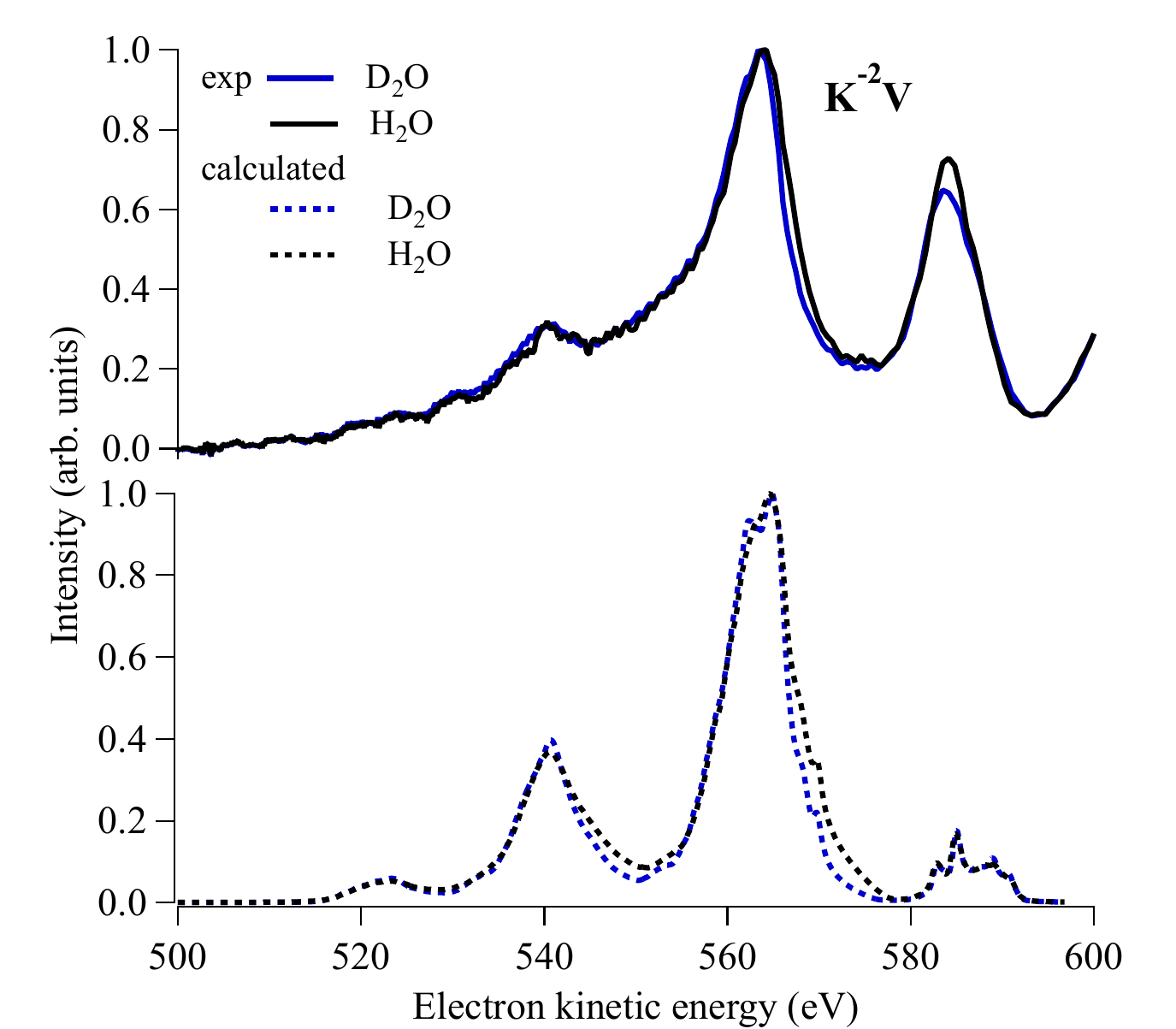}   
 	\caption{Comparison of experimental spectra measured with a photon energy of \unit[617]{eV} (solid lines) with calculated  Auger spectra for a mixture of $\mathrm{K}^{-2}$V states (dashed lines). Black lines show spectra for water, blue lines show spectra for heavy water. The spectra are normalized to have similar peak heights. The experimental spectra show valence photoelectron lines overlapping with the participator Auger contribution located around \unit[585]{eV}. 
	\label{fig:exp2}}
	
\end{figure}

In Fig.~\ref{fig:exp2}, we show the Auger spectra measured at a photon energy of \unit[617]{eV}.
At these photon energies, DCH states are created by a sequence of core-ionization and resonant core-excitation steps, producing singly charged K$^{-2}$V configurations, as discussed in Fig.~\ref{fig:process}.
%%% Ralph's suggestion:
% The spectrum in the energy range of 515 to 575 eV matches well the K-2 spectrum, see Fig.~\ref{fig:exp1}, however, shifted to higher energies by about 7 eV. These spectral features can be explained with spectator Auger decays and the shift can be explained by the extra electron of the K$^{-2}$V configuration as compared to the K$^{-2}$ configuration; this electron leads to a screening effect. In the kinetic energy range above $\unit[580]{eV}$ the calculations predict participator Auger transitions. 
%As can be seen in Fig.~2a of Ref.~\cite{Ismail_PRL}, the participator Auger electrons overlap partly with the photoelectrons of direct valence ionization. Due to the employed photon bandwidth the valence signal is broad and probably also explains the misfit between the calculated and the observed intensity for the spectral feature around 585 eV.
%%% earlier version (things need to be improved in the formulation / merged with Ralph's suggestion?)
The Auger features between 515 and 575 eV appear qualitatively similar to those measured at photon energy of $\unit[750]{eV}$ [Fig.~\ref{fig:exp1}] but are shifted to higher kinetic energies.
%This shift occurs because of the charge screening by the additional electron present in the previously unoccupied orbital (V). 
%% comment regarding this statement: I think we shouldn't put it here. In general it helps when the structure 
%% in the text is 1. describe 2. interpret. And this interpretation comes later (in a much more convincing way) in connection with the calculations
The feature observed at $\unit[585]{eV}$ can be partially attributed to participator Auger transitions involving the additional electron present in the V orbital.
Because of the resonance condition, direct valence photoionization appears at the same photoelectron energy as the participator Auger energies. 
In Fig.~\ref{fig:exp2}, the participator transitions thus overlap with the photoelectron signal from valence ionization of \ce{H2O}$^+$ (K$^{-1}$), as shown in Ref.~\cite{Ismail_PRL}.
Because the employed photon bandwidth is rather broad, it fully covers the participator-Auger contribution above $\unit[585]{eV}$.
%The contribution at $\unit[585]{eV}$ must therefore be attributed to both processes.
Notably, the valence-photoelectron signal is not considered in the theoretical calculations, which explains the difference between the calculated and the observed intensity for this spectral feature.

The experimental spectrum shows a significant difference between water and heavy water in the main Auger peak at $\unit[565]{eV}$ following the same trend as previously observed in Fig.~\ref{fig:exp1}.

The broad photon bandwidth leads to the formation of a mixture of several K$^{-2}$V states.
For the comparison with calculation results, the corresponding calculated Auger spectra are compiled from a mixture that is determined using the configuration interaction calculations outlined in Ref.~\cite{Ismail_PRL}.
These calculations lead us to the following compositions of K$^{-2}$V states: 
$0.24(\mathrm{K^{-2}{4a_1}})
+0.52(\mathrm{K^{-2}{2b_2}})
+0.16(\mathrm{K^{-2}{2b_1}})
+0.08(\mathrm{K^{-2}{5a_1}}$). 
Using this mixture, the computed Auger spectra shows a similar agreement with the experimental spectra as in 
Fig.~\ref{fig:exp1}.
Specifically, the trend showing a more pronounced tail for water than for heavy water is reproduced.
Similar as for Fig.~\ref{fig:exp1}, the lower-energy parts of the Auger spectra are considerably sharper in the calculated spectra as compared with the measured spectra.

The clear isotope effect and the good agreement to the calculated spectra in Figs.~\ref{fig:exp1} and Fig.~\ref{fig:exp2} unequivocally establishes the pronounced effects induced by the rapid proton dynamics on the Auger spectra, confirming previous findings~\cite{marchenko2018ultrafast}.

\begin{figure}
	\includegraphics[angle=0,origin=c,width=0.95\linewidth]{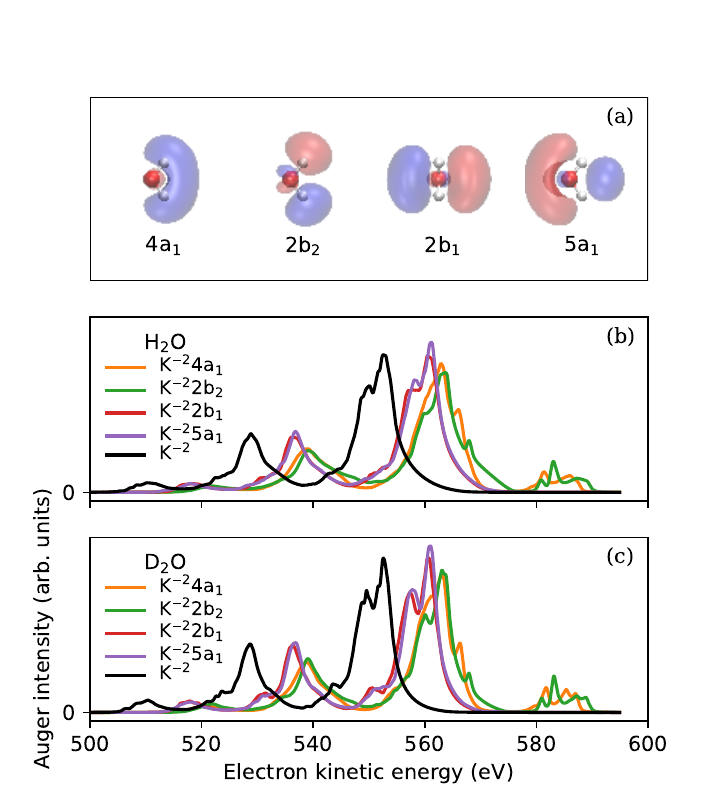}
	\caption{(a) Isosurface plots for the virtual orbitals of the K$^{-2}$ dication. (b) and (c) Calculated Auger spectra for individual K$^{-2}$ and K$^{-2}$V DCH states of \ce{H2O} (b) and \ce{D2O} (c).}
	\label{fig:individualSatellites}
\end{figure}
Figure~\ref{fig:individualSatellites} displays the calculated Auger spectra for the $\mathrm{K}^{-2}$ state and the individual $\mathrm{K}^{-2}$V states. The upper panel [Fig.~\ref{fig:individualSatellites}(a)] shows an isosurface plot of the respective virtual orbitals of the K$^{-2}$ dication that are populated in the respective $\mathrm{K}^{-2}$V states.
The spectra highlight the individual contributions of the resonances towards shaping the high-energy tail of the main peak. 
As one can see, participator contribution above $\unit[580]{eV}$ appear for the $\mathrm{K^{-2}4a_1}$ and the $\mathrm{K^{-2}2b_2}$ configuration. 
For the higher excited $\mathrm{K^{-2}2b_1}$ and $\mathrm{K^{-2}5a_1}$ they are practically absent, because of the diffuse character of the orbitals leading to very small Auger amplitudes for the involved transitions.
The Auger spectra of these two higher-excited DCH states is very similar to the one of the $\mathrm{K}^{-2}$ configuration but shifted by about $\unit[7]{eV}$ to higher energies, indicating that the impact of having an additional electron in the $\mathrm{5a_1}$ or the $\mathrm{2b_1}$ orbital essentially amounts to screening of the ion charge.

The computational results allow us to inspect the dynamics in the K$^{-2}$ state and various K$^{-2}$V states that are responsible for the isotope effect in more details.

\begin{figure}
\includegraphics[angle=0,width=0.95\linewidth]{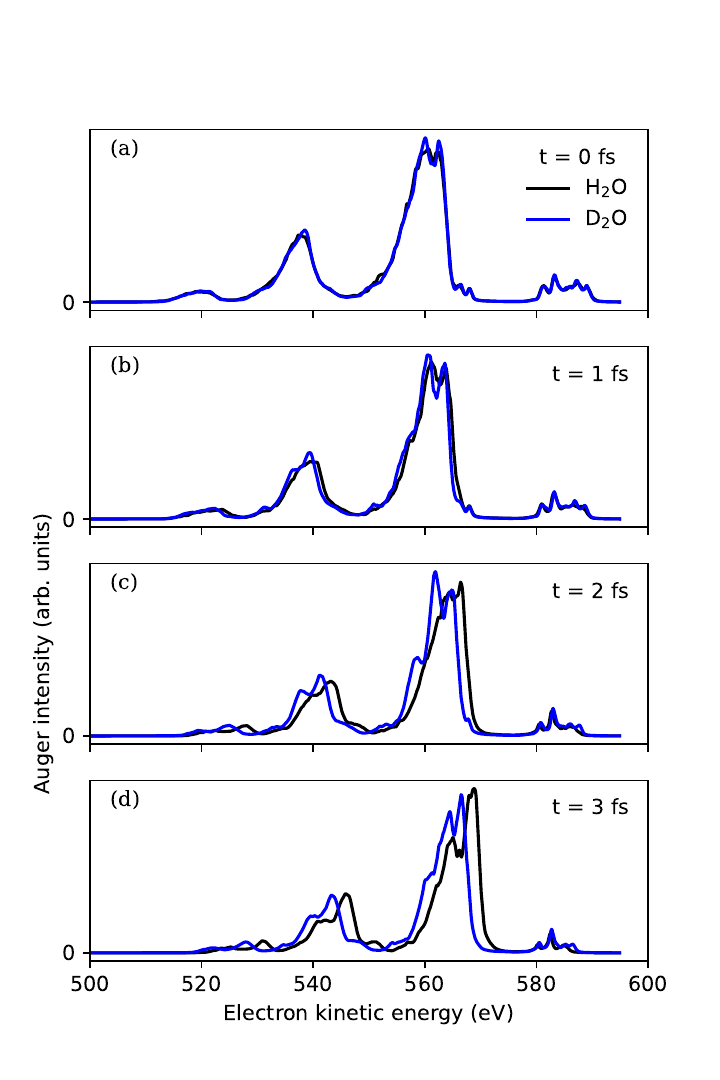}%
	\caption{Instantaneous Auger spectra for selected times after core ionization for an ensemble of K$^{-2}$V states for \ce{H2O} (black) and \ce{D2O} (blue).\label{fig:instantaneous}}
\end{figure}
Figure~\ref{fig:instantaneous} shows calculated spectra from the ensemble of trajectories after selected times of dynamics for the ensemble of $\mathrm{K}^{-2}$V states considered in Fig.~\ref{fig:exp2}.
Spectra for \ce{H2O} as well as for \ce{D2O} are shown.
These instantaneous spectra nicely illustrate how the resulting spectra display effects of core-hole-state dynamics emerging in the marked tail at the high-energy side of the dominant Auger peak. 
%In Fig.~\ref{fig:instantaneous}(a) it is noticeable that small differences between \ce{H2O} and \ce{D2O} already appear at $t=\unit[0]{fs}$ due to the fact that the vibrational wave function in \ce{H2O} is slightly broader than for \ce{D2O}.

As time progress, [Figs.~\ref{fig:instantaneous}(b-d)], the Auger spectrum gradually shifts to higher energies.
This shift is stronger for \ce{H2O} and can be qualitatively understood from the fact that the emission of  protons or deuterons transports charge away from the oxygen atom.
%When the protons have proceeded to a larger distance, 
%{\color{red}creating an additional vacancy in the water molecule \textbf{(the additional vacancy is not very clear to me?  you also  re-discuss this later with only lower Coulomb repulsion which is fine for me!)}}, 
%emitting a valence electron out off the water molecule
At later times, valence electrons thus experience lower Coulomb attraction, which in turn gives rise to faster Auger electrons.
Remarkably, this shift does not occur for the participator contributions above $\unit[580]{eV}$. An observation that can be understood from the specific shape of the participator orbitals as discussed later. 

%The specific shape covering an anti-bonding character along [see~Fig.~\ref{fig:individualSatellites}(a)].
Even though the dynamics can be overall characterized by a rapid symmetric explosion of the molecule, 
the individual satellite states can have a pronounced characteristic core-hole-state dynamics~\cite{marchenko2018ultrafast}.

\begin{figure}
	\includegraphics[angle=0,origin=c,width=\linewidth]{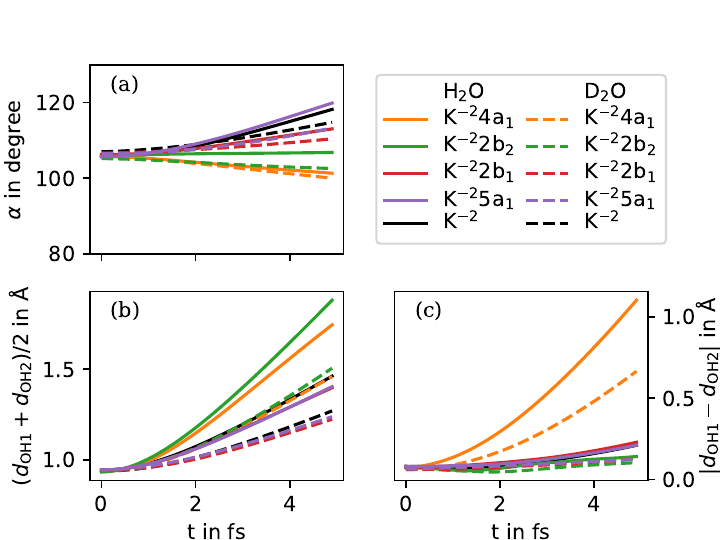}%
 	\caption{Evolution of the internal coordinates (averaged over all trajectories) for individual double-core-hole states. 
  (a) HOH (DOD) bond angle,
  (b) Average oxygen-hydrogen (oxygen-deuterium) distance,
  (c) Absolute difference between oxygen-hydrogen (oxygen-deuterium) distances.
	\label{fig:evolution}}
\end{figure}

Figure~\ref{fig:evolution} shows how the asymmetry, the average OH (OD) bond distance, and the bond angle $\alpha$ evolve as a function of time.
To highlight these differences, the figures show the evolution of trajectories for up to $\unit[5]{fs}$, even though one must keep in mind that the considered double-core-hole configurations have a lifetime of only about $\unit[1.5]{fs}$, and it is thus, according to the exponential decay law, unlikely that such an electronic configuration is present for such a long time.
As one can see, the K$^{-2}$ configuration involves a largely symmetric dissociation into a doubly core-excited oxygen and two protons.
The K$^{-2}$4a$_1$ configuration shows a considerably strong asymmetric fragmentation [Fig.~\ref{fig:evolution}(c)].
In particular, the simulations reveal that this configuration involves asymptotically a fragmentation into a K$^{-2}$V hydroxyl cation (\ce{OH^+}) and a proton. Such a strong antisymmetric character has also been reported before for the single-core-excited $\mathrm{K^{-1}}4a_1$ state~\cite{sankari_2020_nonradiative}.
Moreover, the K$^{-2}$ and K$^{-2}$5a$_1$ configurations show a trend for HOH angle opening dynamics, whereas the opposite (angle closing) can be seen for the K$^{-2}$4a$_1$ configuration [Fig.~\ref{fig:evolution}(a)].
The configuration K$^{-2}$2b$_2$ stands out by exhibiting a particularly rapid symmetric fragmentation [Fig.~\ref{fig:evolution}(b)].
Overall, one can see that the configurations K$^{-2}$4a$_1$ and K$^{-2}$2b$_2$ induce a somewhat more rapid fragmentation than the bare K$^{-2}$ configuration, whereas higher excited states like K$^{-2}$2b$_1$ and K$^{-2}$5a$_1$ show a fragmentation dynamics similar to the K$^{-2}$ configuration.
The variations between the K$^{-2}$V states can be understood from the influence of the electron in the previously unoccupied orbital [see Fig.~\ref{fig:individualSatellites}(a)]. 
The antibonding character of the 4a$_1$ and the 2b$_2$ orbital along the OH bonds pushes the protons more rapidly away.
In addition, the 4a$_1$ orbital has some bonding character between the hydrogen atoms and thus induces lowering of the HOH angle.

The antibonding character of the orbitals 4a$_1$ and 2b$_2$ also explains the distinct behavior of the participator lines along the dynamics observed in Fig.~\ref{fig:instantaneous}.
Whereas spectator Auger contributions become faster in energy as the dissociation proceeds because the involved valence orbitals become less bound as the Coulomb attraction to the protons reduces with increased OH bond length, the participator contributions are, in addition, impacted by the respective orbital's antibonding character along the OH bond, making them \textit{stronger} bound as the dissociation proceeds.
As a result, participator contributions do not shift that strongly in the instantaneous Auger spectra shown in Fig.~\ref{fig:instantaneous}.

%Moreover, this effect is notably amplified in the more dissociative K$^{-2}$V state, where the nuclei can exhibit more substantial motion within the same timeframe compared to the K$^{-2}$ state, resulting in a more significant difference in the HS Auger spectra. This observation  not only serves as a double confirmation of the ultrafast motion but also importantly confirms the theoretically predicted dependence of the dynamic response on the populated DCH states. Additionally, it highlights the ability to control these dynamics within the subfemtosecond scale through photon energy tuning. 

\section{Conclusions} \label{sec:Conclusions}
We have experimentally and theoretically investigated double-core-hole states in isolated water and heavy water molecules generated through the sequential absorption of two x-ray photons. 
Theoretical calculations of the Auger process and the involved core-hole-state dynamics successfully reproduce the experimental data, enabling an in-depth examination of the dynamics in K$^{-2}$ and K$^{-2}$V states responsible for the observed isotope effect. 
Our measurements confirm earlier synchrotron measurements, where DCH states were created via one-photon absorption~\cite{marchenko2018ultrafast}.
We note that analogous observations have very recently also been made for the one-photon-DCH-Auger spectrum of liquid water~\cite{trinter_2024_radiationless} demonstrating a considerable 
impact of the liquid environment on the DCH-Auger electron energy.
The comparison of the Auger spectra obtained from the two isotopic systems in the gas-phase provides an unequivocal confirmation of the ultrafast proton motion during the lifetime of the DCH states. 
By selectively populating different K$^{-2}$ and K$^{-2}$V states, Auger spectra of both species could be disentangled and distinct dynamic behaviors were observed.

\begin{acknowledgements}
I.I. and L.I. contributed equally to this work.

We acknowledge European XFEL in Schenefeld, Germany, for the provision of XFEL beam time at the SQS instrument and would like to thank the staff for their assistance. L.I. acknowledges support from DESY (Hamburg, Germany), a member of the Helmholtz Association HGF, and the scientific exchange and support of the Centre for Molecular Water Science (CMWS). A.F., R.T., and S.C. thank Labex MiChem part of French state funds, and managed by the ANR within the Investissements d’Avenir programme (Sorbonne Université, ANR-11-IDEX-0004-02), for providing A.F.'s PhD funding. D.P., A.V., J.D.B. and M.S. acknowledge the financial support of the CNRS, and GotoXFEL program. T.M and N.V. acknowledge funding from the European Union's Horizon 2020 research and innovation programme under the Marie Skłodowska-Curie grant agreement No [860553]. F.T. acknowledges funding by the Deutsche Forschungsgemeinschaft (DFG, German Research Foundation) - Project 509471550, Emmy Noether Programme and acknowledges support by the MaxWater initiative of the Max-Planck-Gesellschaft. Work by D.K., K.L., G.D. and L.Y.  was supported by the U.S. Department of Energy, Office of Science, Basic Energy Science, Chemical Sciences, Geosciences and Biosciences Division under Contract No. DE-AC02-06CH11357. M.M. acknowledges support by the DFG, German Research Foundation – SFB-925 – project 170620586 and by the Cluster of Excellence Advanced Imaging of Matter of the DFG, EXC 2056, Project ID 390715994.
\end{acknowledgements}

The experimental data were collected during user beamtime 2620. The metadata are available at https://doi.org/10.22003/XFEL.EU-DATA-002620-00.

%\nocite{*}
%\bibliographystyle{apsrev4-2}
%\input{main-m.bbl}
\bibliography{refs}% Produces the bibliography via BibTeX.

\end{document}